\documentclass[]{spie}  

\usepackage[]{graphicx}

\title{Stripes and superconductivity in cuprate superconductors} 

\author{J. M. Tranquada
\skiplinehalf
Physics Dept., Brookhaven National Laboratory, Upton, NY 11973, USA
}

\begin{document} 
  \maketitle 

\begin{abstract}
One type of order that has been observed to compete with
superconductivity in cuprates involves alternating charge and
antiferromagnetic stripes.  Recent neutron scattering studies indicate
that the magnetic excitation spectrum of a stripe-ordered sample is very
similar to that observed in superconducting samples.  In fact, it now
appears that there may be a universal magnetic spectrum for the
cuprates.  One likely implication of this universal spectrum is that
stripes of a dynamic form are present in the superconducting samples.  On
cooling through the superconducting transition temperature, a gap opens
in the magnetic spectrum, and the weight lost at low energy piles up
above the gap; the transition temperature is correlated with the size of
the spin gap.  Depending on the magnitude of the spin gap with respect
to the magnetic spectrum, the enhanced magnetic scattering at low
temperature can be either commensurate or incommensurate.  Connections
between stripe correlations and superconductivity are discussed. 
\end{abstract}



\section{Introduction}

\begin{figure}
   \begin{center}
   \begin{tabular}{c}
   \includegraphics[height=4cm]{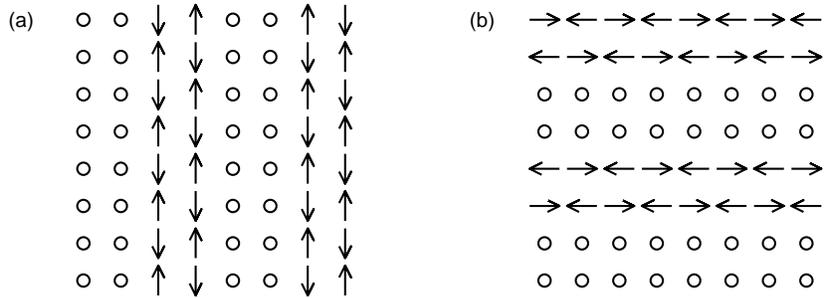}
   \end{tabular}
   \end{center}
   \caption[example] 
   { \label{fig:stripes} 
Cartoon of two equivalent domains of stripe order in CuO$_2$ planes: (a)
vertical stripes, (b) horizontal stripes.  Only Cu sites are indicated,
with arrows indicating ordered magnetic moments and circles
indicating hole-rich charge stripes.  The absolute phase of the stripe
order with respect to the lattice has not been established by
experiment, so stripes could also be centered on Cu sites, rather than
on bonds.} 
   \end{figure}

The concept of charge stripes\cite{kive03,zaan01,mach89} is a
controversial one in the field of high-temperature superconductivity. 
There is direct evidence from neutron diffraction measurements for charge
and spin stripe order in a couple of cuprates,\cite{tran95a,fuji04}
La$_{1.6-x}$Nd$_{0.4}$Sr$_x$CuO$_4$ and La$_{2-x}$Ba$_x$CuO$_4$ with
$x\approx\frac18$; however, the ordering of stripes is correlated with
the depression of the superconducting transition temperature, $T_{\rm
c}$.\cite{ichi01} (For reference, Fig.~\ref{fig:stripes} shows examples
of possible stripe domains.) While it is clear that static {\it ordering}
of stripes competes with superconductivity, this does not mean that
dynamic stripes are necessarily bad for superconductivity.  In this
paper, I will argue the case that stripes, in fact, underlie the
high-temperature superconductivity in under- to optimally-doped cuprates
(with the implication that stripes, or more general forms of charge
inhomogeneity, are essential to the superconductivity).

\section{Stripe order is common}

As a starting point to understanding the cuprates, it makes sense to
consider the behavior of related compounds.  Consider the systems
La$_{2-x}$Sr$_x$$M$O$_4$, where, of course, the compounds with $M=$ Cu are
superconducting for $0.05 < x < 0.25$.  For $M=$ Ni, Co, and Mn, the
trend is to have either charge and locally-antiferromagnetic spin order,
or else a metallic ferromagnetic state.  In the case of $M=$ Ni, diagonal
charge and spin stripes are observed over the
range\cite{sach95,tran96a,yosh00} $0.2 < x < 0.5$, with checkerboard
order\cite{kaji03} at $x=0.5$; dynamic stripe-like spin correlations
survive\cite{lee02,bour03} in the disordered state of
La$_{2-x}$Sr$_x$NiO$_4$.  A state close to checkerboard order is also
observed\cite{zali00} for $M=$ Co and
$x=0.5$.  When $M=$ Mn and
$x=0.5$, the presence of several degenerate, partially-filled $d$ levels
leads to a complicated phase at low temperature involving charge,
orbital, and spin ordering.\cite{ster96}  The ferromagnetic metallic
state is observed in pseudo-cubic manganese perovskites.\cite{schi95}

The point here is that there is a common tendency in layered
transition-metal oxides for the holes doped into the two-dimensional (2D)
antiferromagnetic planes to segregate, order, and coexist with locally
antiferromagnetic domains.  In three-dimensional perovskites, metallic
behavior tends to be coupled with ferromagnetism.  Thus, the cuprates are
surprising not only for their superconductivity, but also for their
metallic normal state with antiferromagnetic spin excitations. 
Nevertheless, in seeking to understand these properties, it should not be
surprising, given the context, that charge segregation effects may be
relevant.

\section{Universal magnetic excitation spectrum}

\begin{figure}
   \begin{center}
   \begin{tabular}{c}
   \includegraphics[height=10cm]{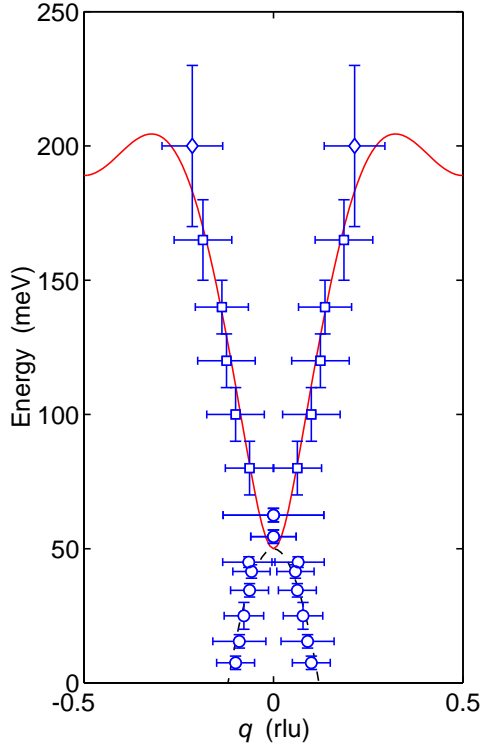}
   \end{tabular}
   \end{center}
   \caption[example] 
   { \label{fig:disp} 
Symbols: experimentally measured dispersion\protect\cite{tran04} of
magnetic excitations along ${\bf Q}=(0.5+q,0.5,l)$ in stripe-ordered
La$_{1.875}$Ba$_{0.125}$CuO$_4$.  The solid line is the
calculated\protect\cite{barn94} dispersion along a two-leg ladder with
$J=100$~meV.} 
   \end{figure}

Considerable progress has been made in the last couple of years in
characterizing the magnetic excitation spectra of superconducting
cuprates with inelastic neutron scattering.  In the case of
YBa$_2$Cu$_3$O$_{6+x}$, it has become clear that, besides the so-called
``resonance'' peak, there are also excitations that disperse upwards to
higher energies and downwards to lower
energies.\cite{arai99,bour00,hayd04,rezn04,stoc04,pail04}  A similar
``hour-glass'' dispersion is found over a range of $x$, from 0.5 to
0.95.  A very similar dispersion has also been observed in
La$_{1.875}$Ba$_{0.125}$CuO$_4$ (LBCO)\cite{tran04} and in
optimally-doped La$_{2-x}$Sr$_x$CuO$_4$;\cite{chri04} a plot of the
dispersion for LBCO is shown in Fig.~\ref{fig:disp}.  As the LBCO sample
exhibits stripe order\cite{fuji04} (see Fig.~\ref{fig:stripes}), it is
natural to interpret the magnetic excitations as spin waves of the
ordered configuration.  It turns out that the dispersion is not
consistent with linear spin-wave theory,\cite{krug03,carl04} in that we
do not observe symmetric dispersion (and intensity) about the
incommensurate ordering wave vectors.  (In contrast, linear spin-wave
theory gives a good description of the spin excitations in diagonally
stripe-ordered La$_{2-x}$Sr$_x$NiO$_4$.\cite{bour03,boot03,woo05})

Another way to think about the spectrum is to take into account that
bond-centered stripes at $x=\frac18$ define two-leg spin ladders (see
Fig.~\ref{fig:stripes}).  The dispersion of triplet excitations within an
isolated ladder\cite{barn94} with $J=100$~meV is shown by the solid line
in Fig.~\ref{fig:disp}; it agress with the measurements surprisingly
well.  To account for the downward dispersion, one must allow for a
coupling between the ladders (across the charge
stripes).\cite{vojt04,uhri04}  An alternative approach, involving
calculating excitations with respect to a particular mean-field stripe
state,\cite{seib05} also gives good agreement with the data.  In
contrast, it is not possible to reproduce the full spectrum with a model
based on checkerboard order.\cite{vojt05}

While the simple spin-only models do remarkably well, they have
shortcomings.  In particular, they do not properly describe the
anisotropic magnetic scattering measured in a detwinned sample of
YBa$_2$Cu$_3$O$_{6.85}$ by Hinkov {\it et al.}\cite{hink04}   Of course,
that sample has no static stripe order.  The results may be compatible
with dynamic stripes.

In LBCO, the magnetic spectrum at low energies ($<12$~meV) is not very
sensitive to the presence of stripe order,\cite{fuji04} although the
frequency dependence of the scattered intensity is.  Measurements are
underway to check for temperature dependence at energies up to 100 meV. 
The similarities of the excitations in the ordered and disordered states
suggests that the stripe correlations survive in a dynamic form in the
disordered state.  The similarities among the different cuprate families
further suggest that dynamic stripes are common in these materials and
coexist with superconductivity.

\begin{figure}
   \begin{center}
   \begin{tabular}{cc}
   \includegraphics[height=3.5cm]{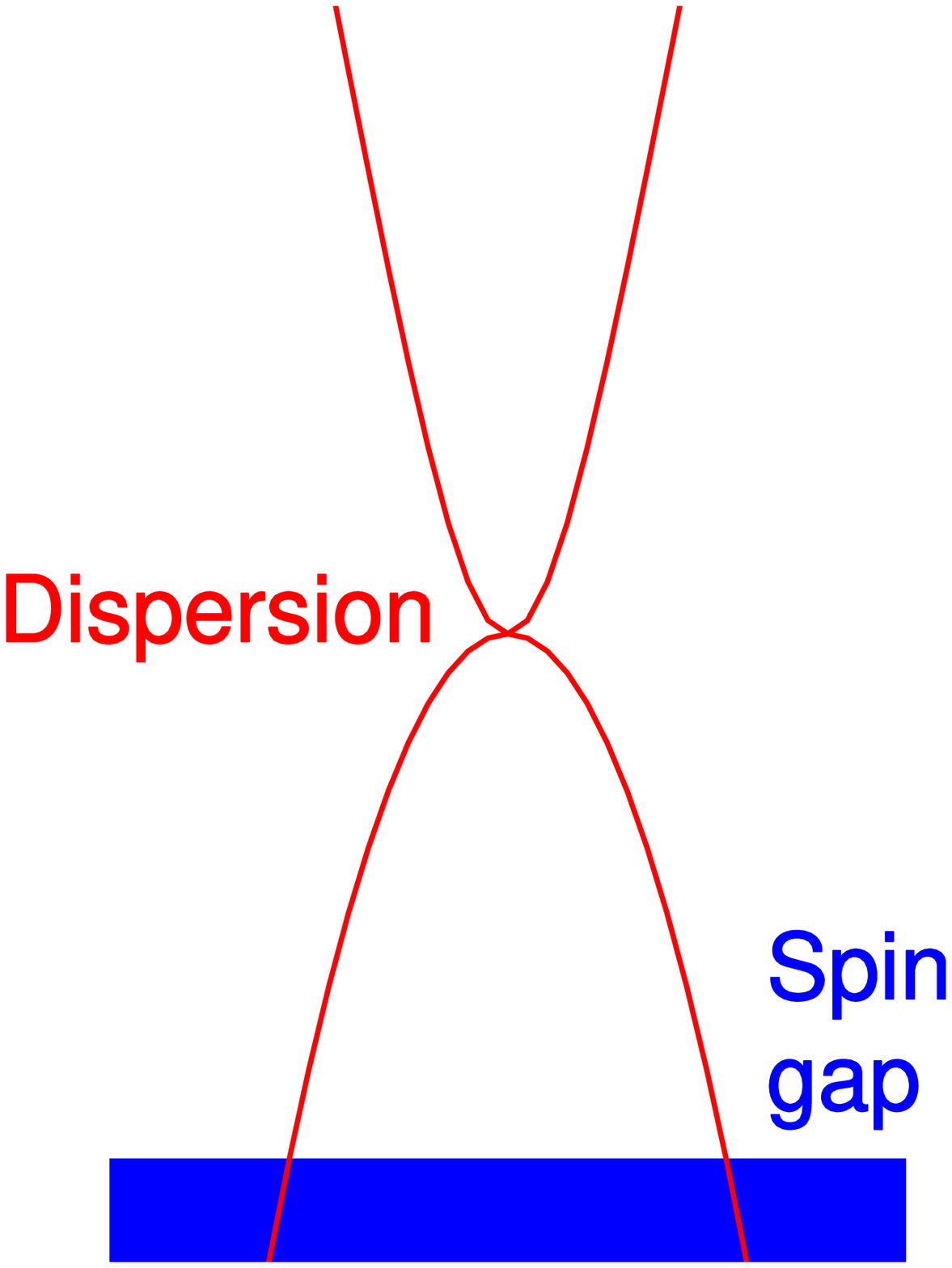} \hspace{20pt}
   \includegraphics[height=3.5cm]{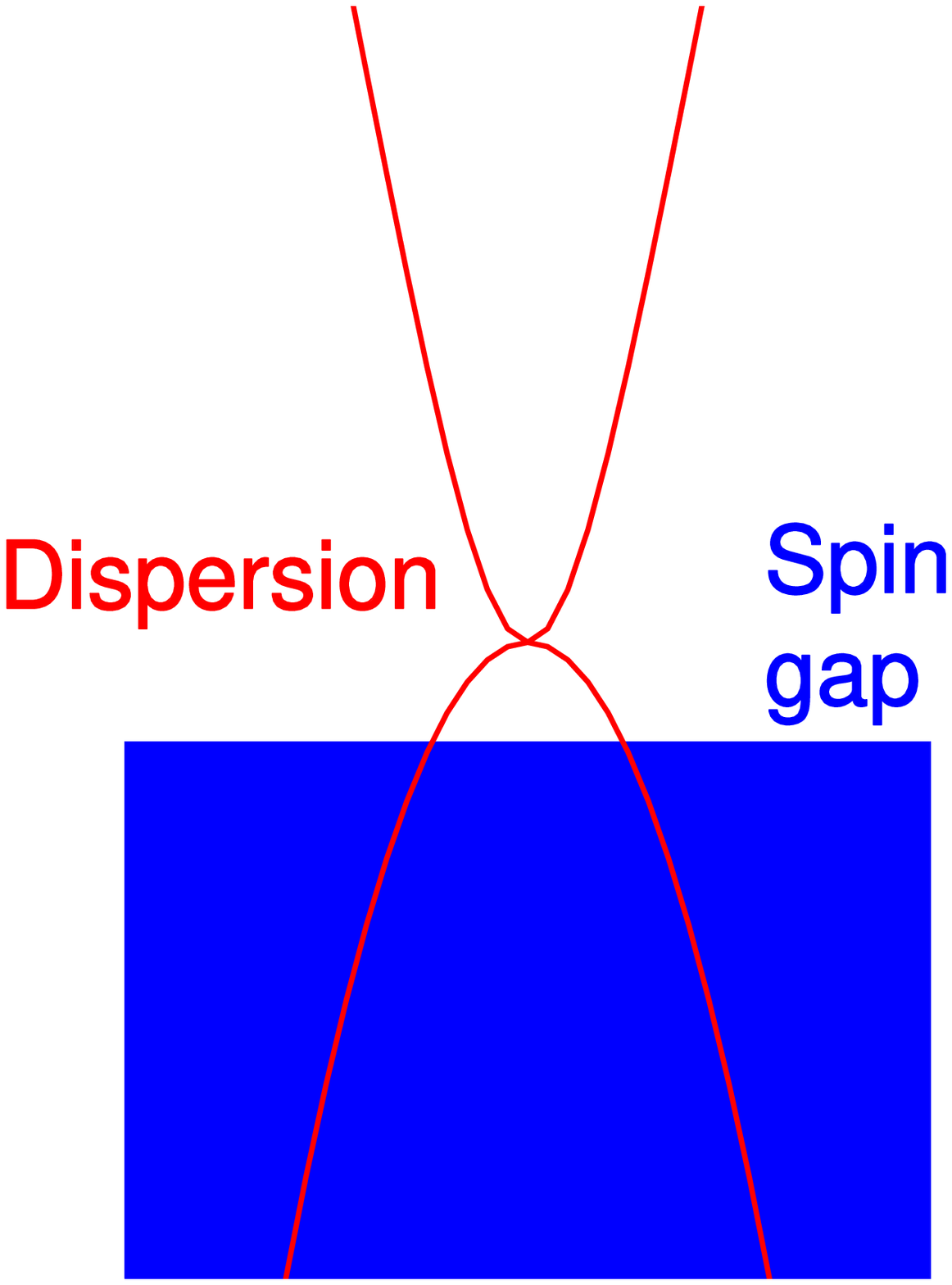}
   \end{tabular}
   \end{center}
   \begin{center}
   \begin{tabular}{c}
   \includegraphics[height=5.5cm]{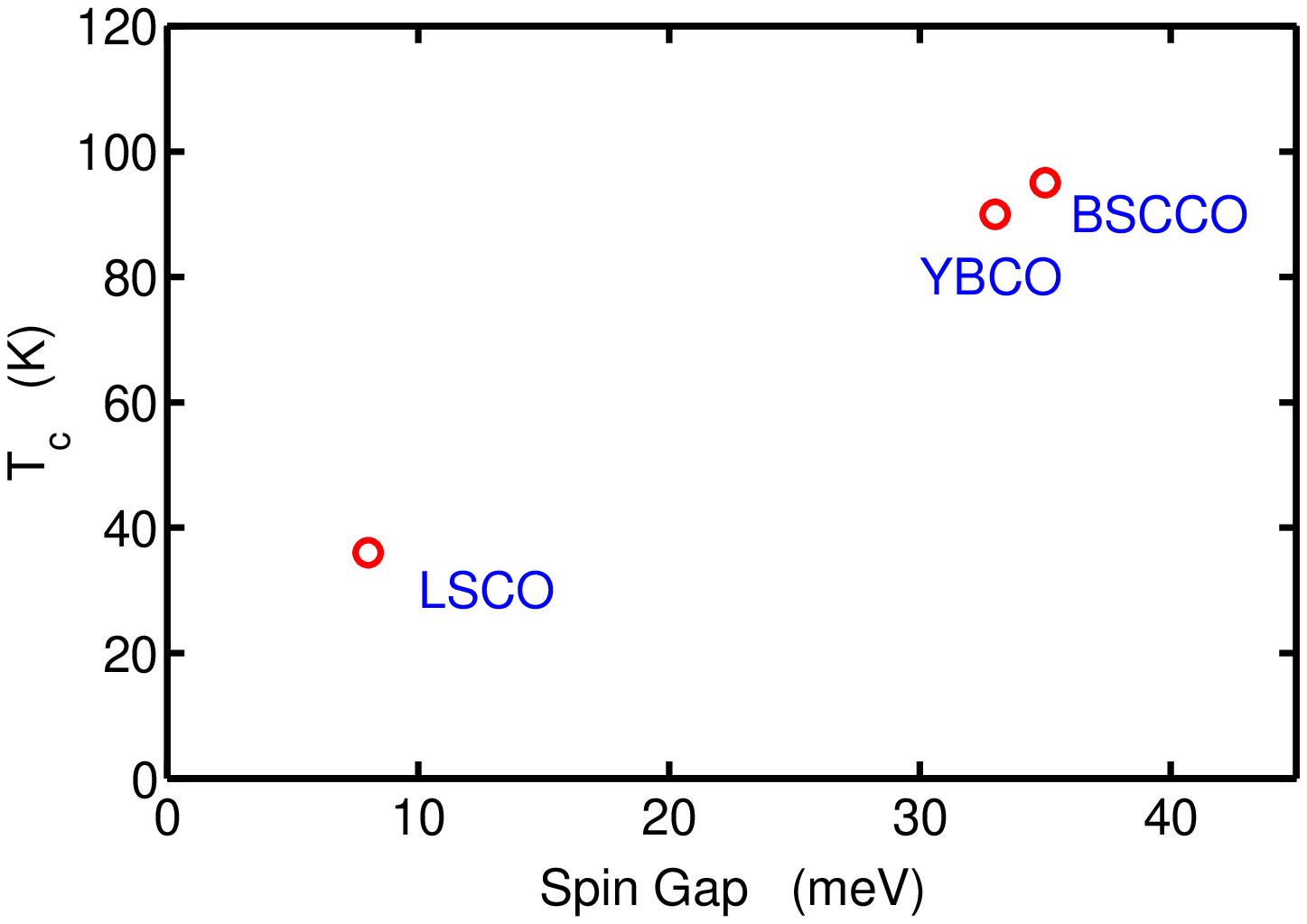}
   \end{tabular}
   \end{center}
   \caption[example] 
   { \label{fig:spin_gap} 
Top left: schematic diagram of the magnetic dispersion and the spin gap
for a system such as La$_{2-x}$Sr$_x$CuO$_4$ (LSCO), where the magnitude
of the spin gap is much smaller than the saddle point.  Top right:
diagram for optimally doped YBa$_2$Cu$_3$O$_{6+x}$ (YBCO) and
Bi$_2$Sr$_2$CaCu$_2$O$_{8+\delta}$ (BSCCO), where the spin gap is just a
bit smaller than the saddle point energy.  Bottom: plot of $T_{\rm c}$
vs.\ spin gap energy for three optimally-doped cuprates.} 
   \end{figure}

In optimally-doped superconducting samples, an energy gap appears in the
magnetic excitations for temperatures below $T_{\rm c}$.  The
measurements of Christensen {\it et al.}\cite{chri04} on
La$_{1.84}$Sr$_{0.16}$CuO$_4$ indicate that the weight below the spin gap
is shifted to energies just above the spin gap.  In this particular case,
where the spin gap is small ($\sim8$ meV, depending on how one measures
it) compared to the saddle-point energy of the dispersion, the enhanced
intensity below $T_{\rm c}$ all occurs at incommensurate wave vectors
(see Fig.~\ref{fig:spin_gap}).  Applying a magnetic field tends to shift
some of this intensity back into the gap.\cite{tran04b,lake01}  In
compounds such as YBa$_2$Cu$_3$O$_{6+x}$ and
Bi$_2$Sr$_2$CaCu$_2$O$_{8+\delta}$, where the spin gap is much larger and
closer to the saddle-point energy, much of the transferred scattering
weight below $T_{\rm c}$ appears at the saddle point, yielding the
commensurate resonance.\cite{bour00,dai01,fong99}  The saddle point
energy (at optimum doping) appears to scale with the superexchange
energy (which is somewhat larger in La$_2$CuO$_4$ than in
YBa$_2$Cu$_3$O$_6$).  The spin-gap energy varies substantially among
different cuprate families, and shows a correlation with $T_c$, as
indicated in Fig.~\ref{fig:spin_gap}.  The idea that a universal magnetic
spectrum plus a spin gap might explain various observations in the
cuprates was first suggested by Batista {\it et al.}\cite{bati01}

\section{Stripes and superconductivity}

I have argued that the magnetic excitation spectrum observed in the
cuprates is associated with stripe correlations.  If this is correct, and
if magnetic correlations are important to the mechanism of
superconductivity, then it appears that dynamic stripes may not only
underlie the superconductivity in the hole-doped cuprates, but also be
an essential component of the superconductivity.  One proposed mechanism
for the superconductivity, the spin-gap proximity effect,\cite{emer97} is
based on this sort of picture.  The correlation between $T_c$ and spin
gap energy shown in Fig.~\ref{fig:spin_gap} is predicted by this
approach.\cite{carl03}  In fact, it has been argued that charge
inhomogeneity is essential to obtaining high-temperature
superconductivity.\cite{arri04}

\begin{figure}
   \begin{center}
   \begin{tabular}{c}
   \includegraphics[height=10cm]{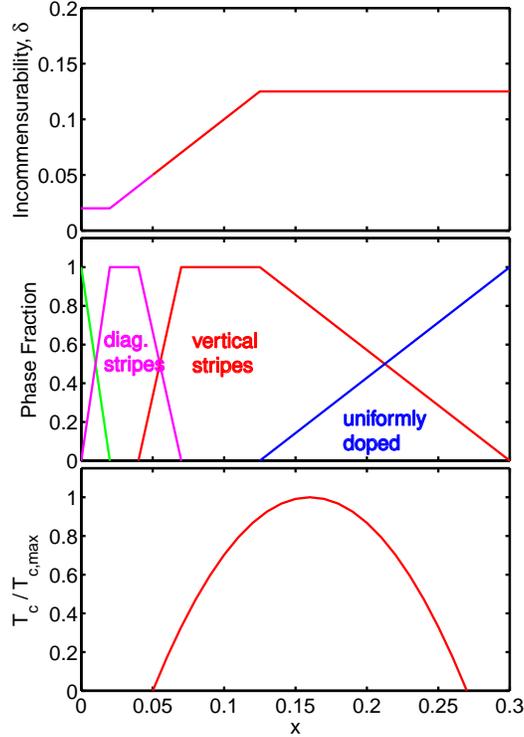}
   \end{tabular}
   \end{center}
   \caption[example] 
   { \label{fig:stripe} 
Top: schematic diagram of magnetic incommensurability (elastic or
low-energy inelastic) measured in La$_{2-x}$Sr$_x$CuO$_4$ by neutron
scattering as a function of doping.\protect\cite{yama98a,mats00b,mats02}
Middle: schematic diagram suggesting regions of existence and coexistence
of various electronic states; from left, antiferromagnetic order, diagonal
stripes, vertical stripes, uniformly-doped phase.  Bottom: typical curve
of $T_{\rm c}$ vs.\ doping.} 
   \end{figure}

There are, of course, limits to the stripe picture.  We know from the
work of Yamada {\it et al.}\cite{yama98a,mats00b} on LSCO that the
magnetic incommensurability, which is inversely proportional to the
stripe spacing, increases linearly from $x=0.02$ to about 0.13,
saturating beyond that point (see top panel of Fig.~\ref{fig:stripe}). 
(Similar behavior has been reported for YBCO,\cite{dai01} although it
appeared to saturate at a smaller incommensurability.  This is due to the
fact that the incommensurability was measured at about 30 meV, and did
not take into account the dispersion of the excitations to larger wave
vectors at lower energies.)  In the saturated region, where the charge
stripes are separated by about 4 lattice spacings, it seems unlikely that
added holes would be forced into the existing stripes, increasing the hole
density per stripe.  It seems more likely that that the added holes will
form uniformly-doped regions, so that, with increasing $x$, the fractional
area occupied by stripes at any point in time will decrease.  This is
indicated schematically in the middle panel of Fig.~\ref{fig:stripe}. 

Wakimoto {\it et al.}\cite{waki04} have recently shown that the
low-energy ($\sim6$~meV) magnetic scattering decreases rapidly with
overdoping beyond $x\sim0.2$, going to zero by $x=0.30$.   In the stripe 
picture, the magnetic excitations are associated with the
stripe-correlated regions, so that the experimental results are
consistent with a decrease in the stripe fraction in the overdoped
regime.  Measurements are currently underway to test whether this
decrease in magnetic signal also applies at higher energies, up to 100
meV.

\acknowledgments     
 
The perspective expressed in this paper is strongly influenced by
frequent discussions with Steve Kivelson and Eduardo Fradkin.  Work at
Brookhaven is supported by the Office of Science, U.S. Department of
Energy, under Contract No.\ DE-AC02-98CH10886.



\end{document}